\documentstyle[psfig,emulateapj]{article}
\def\lcdm  {\rm $\Lambda$CDM\ }
\def\ldcdm  {\rm $\Lambda$DCDM }
\def\etal   {{et~al.}\ }

\def\kpc{{\rm\,kpc}}

\def\vol#1  {{{#1}{\rm,}\ }}

\def\etal{et al.\ }

\newcount\refno
\newcount\rfno
\def\eq{$^{\the\refno\ }$\advance\refno by 1}
\def\ad{\advance\rfno by 1}

\def\clock{\count0=\time \divide\count0 by 60
     \count1=\count0 \multiply\count1 by -60 \advance\count1 by \time
     \number\count0:\ifnum\count1<10{0\number\count1}\else\number\count1\fi}

\begin{document}
\title{Decaying Cold Dark Matter Model and Small-Scale Power}
\author{Renyue Cen\altaffilmark{1}}
\altaffiltext{1} {Princeton University Observatory, Princeton University, Princeton, NJ 08544; cen@astro.princeton.edu}

\begin{abstract}

The canonical cosmological constant dominated cold dark matter model 
(\lcdm)
may possess too much power on small scales at $z=0$,
manifested as central over-concentration of dark matter 
and over-abundance of dwarf galaxies.
We suggest an alternative model, $\Lambda$DCDM, 
where one half of the cold dark matter
particles decay into relativistic particles by $z=0$.
The model successfully
lowers the concentration of dark matter in dwarf galaxies as well
as in large galaxies like our own {\it at low redshift},
while simultaneously retaining the virtues of the \lcdm model.
The model solves the problem of
over-production of small dwarf galaxies in the \lcdm
{\it not by removing them but by 
identifying them with failed, ``dark" galaxies},
where star-formation is quenched due to dark matter evaporation
and consequent halo expansion.
A dramatic difference between the \ldcdm model 
and other proposed variants of the \lcdm model
is that the small-scale power {\it at high redshift} ($z>2$)
in the \ldcdm model is enhanced
compared to the \lcdm model.

A COBE-and-cluster normalized \ldcdm model
can be constructed with the following parameters:
$H_0=60$km/sec/Mpc,
$\lambda_0=0.60$, $\Omega_{0,CDM}=0.234$,
$\Omega_{0,b}=0.044$,
$n=1.0$, and $\sigma_8=1.06$.
A clean test of this model can be made by measuring
the evolution of gas fraction in clusters.
The prediction is that the gas fraction should decrease
with redshift and is smaller by $31\%$ at $z=1$ than
at $z=0$. X-ray and Sunyaev-Zel'dovich effect observations
should provide such a test.

\end{abstract}


\section{Introduction}

While the canonical \lcdm model
is remarkably successful in many ways
(Ostriker \& Steinhardt 1995; Bahcall \etal 1999),
there is now some tentative evidence that
it may have too much power on small scales ($l\sim 1-100$\kpc) today.
Evidence includes a large excess 
of dwarf galaxies
(Klypin \etal 1999; Moore \etal 1999),
the over-concentration of dark matter in 
dwarf galaxies 
(Moore 1994; Flores \& Primack 1994;
Burkert 1995;
McGaugh \& de Blok 1998;
Moore \etal 1999)
as well as
in 
large galaxies 
(Navarro \& Steinmetz 2000).
Without impairing the many notable virtues of the \lcdm model,
in this {\it Letter} 
we suggest a \lcdm model
in which one half of the CDM particles
decay into relativistic particles by $z=0$.
Decaying CDM was
suggested before in attempts to
save the CDM model 
in an Einstein-de Sitter universe 
(Turner, Steigman, \& Krauss 1984;
Doroshkevich \& Khlopov 1984;
Olive, Seckel, \& Vishniac 1985).
Suggestions were also made 
in the context of neutrino models 
for similar rescue missions
(Davis \etal 1981; Hut \& White 1984).

%

\section{Decaying Cold Dark Matter Model}

To illustrate the effect of CDM decay
let us examine how the profile of a halo would be altered.
Suppose a pure CDM halo, 
formed at some high redshift $z_{halo}$,
has an NFW (Navarro, Frenk, \& White 1997) density profile 
\begin{equation}
{\rho (r)\over \rho_{crit}} = {\delta_c\over (r/r_c)(1+r/r_c)^2},
\end{equation}
\noindent 
with an initial concentration parameter $c_i\equiv r_{200,i}/r_{c,i}$,
where $r_{200}$ is the radius within which
the mean density is $200\rho_{crit}$ ($\rho_{crit}$ is the critical density),
$r_c$ is the characteristic ``core" radius 
and $\delta_c$ is the characteristic (dimensionless) density.
Suppose that a fraction, $1-y$, of the CDM particles
will decay by $z=0$.
Since the proposed decay lifetime $\tau$ is greater than $t_0$ 
(the current age of the universe)
and the orbital periods of particles inside $r_{200}$
for halos of interest, 
one may assume that the change in the potential 
of the halo due to CDM decay is gradual
resulting in an adiabatic expansion of the halo.
For this illustrative example we also assume
that the halo maintains the NFW profile during expansion.
We identify the ``initial" (subscript ``i") halo configuration
with the halo configuration at $z=0$ in the canonical
\lcdm model,
where CDM particles would have been stable, 
and the ``final" (subscript ``f") halo configuration
with the halo configuration at $z=0$ in the \ldcdm model.
Using virial theorem
it can be shown 
that a particle at initial radius $r_i$ 
will move out to $r_f$: $r_f = r_i/y$.
The final ``core" radius is therefore
\begin{equation}
r_{c,f} = r_{c,i}/y.
\end{equation}
\noindent 
The mass inside $r_{200,i}/y$
is $yM_{200,i}$ by $z=0$,
where 
$M_{200,i}$ is the initial mass within $r_{200,i}$,
and the final density
within $r_{200,i}/y$ is $y^4 \rho_{200}$.
Defining $-\alpha$ as the effective
slope of the density profile at $\sim r_{200}$,
we obtain $r_{200}$ 
approximately 
\begin{equation}
r_{200,f} \approx y^{4/\alpha-1}r_{200,i}.
\end{equation}
\noindent 
Another way to obtain $r_{200,f}$ is to use
equation (1) directly to solve for $r^\prime$ 
within which the initial density is $200y^{-4}\rho_{crit}$.
The resulting equation is
$w(1+c_i w)^2 = y^4 (1+c_i)^2$, where $w=r^\prime/r_{200,i}$.
Using $y=0.5$ and $c_i=30$ one obtains
$w=0.384$, resulting in the final virial radius
$r_{200,f}=0.384y^{-1}r_{200,i}=0.77r_{200,i}$ (for $y=0.5$),
which is what equation (3) gives with $\alpha=2.9$
(consistent with the known slope of halos near $r_{200}$).
For our present purpose, 
we will simply adopt $\alpha=3$ and use the analytic form of equation (3)
for subsequent analyses.
Combining equations (2,3) yields,
\begin{equation}
c_{f} =  y^{4/\alpha} c_{i}.
\end{equation}
\noindent 
For $y=0.5$, it says $c_i/c_f=2.5$. 
The circular velocity due to the CDM halo is  (Navarro \etal 1997)
\begin{equation}
\left [{V_c(r)\over V_{200}}\right]^2 = {1\over x} {\ln (1+cx)-(cx)/(1+cx)\over \ln (1+c)-c/(1+c)},
\end{equation}
\noindent 
where $x\equiv r/r_{200}=r/cr_c$.
$V_{200,f}$ and $V_{200,i}$ are related:
\begin{equation}
V_{200,f} = y^{4/\alpha -1} V_{200,i}
\end{equation}
\noindent 
Figure (1) shows
rotation curves 
and mass  profiles 
for the initial 
and final 
halos (with $y=0.5$).
The reduction in $V_{max}$, $c$ and mass of the halo,
and the increase in $r_{max}$ 
in the \ldcdm model 
should alleviate the density profile/concentration crisis.
For example, the CDM mass within the solar circle 
($\sim 0.1r_{200}$) will be reduced by a factor about $3.5$,
as seen in Figure 1.
But for Milky Way like galaxies where CDM
mass is not dominant today within the relevant radius,
the reduction should be smaller, likely 
in the range $2-3$, which will bring
the model into agreement with observations (Navarro \& Steinmetz 2000).
Detailed simulations including other important effects
such as mergers and tidal fields
should provide a more precise anwser.
The effect of the decay of CDM particles
on the abundance of small dwarf galaxies 
is also favorable.
This, and additional attractive features 
as well as potential tests of the model,
will be discussed in the next section.

\section{Discussion}

\subsection{A Fiducial \ldcdm Model}

Let us construct a fiducial \ldcdm model 
with the following parameters:
$\Omega_{0,CDM}=0.234$ (CDM density at $z=0$),
$\lambda_0=0.60$, $\Omega_{0,b}=0.044$ (baryonic density at $z=0$),
$\Omega_{0,r}=0.122$ (relativistic matter density at $z=0$) and
$h\equiv H_0/100$km/sec/Mpc$=0.60$.  
The assumption that one half of the CDM particles 
decay by $z=0$
translates into an exponential decay lifetime 
$\tau\sim 1.44t_0$, where $t_0=0.84H_0^{-1}$ 
is the current age of the universe (equal to $13.9$ Gyrs for $h=0.60$).
At $z>>1$ the total non-relativistic matter 
density is
$\rho_{i,NR}=(\Omega_{0,b} + 2\Omega_{0,CDM})(1+z)^3\rho_{crit}=0.512(1+z)^3\rho_{crit}$,
where $\rho_{crit}$ is the critical density at $z=0$.
Eke \etal (1996) cluster normalization
requires $\sigma_8=(0.52\pm 0.04)\Omega_{0,NR}^{-0.52+0.13\Omega_{0,NR}}$
for $\lambda_0+\Omega_{0,M}=1$,
which gives $\sigma_8=1.01\pm 0.08$.
Strictly speaking this fitting formula does not
apply to \ldcdm models.
But we expect the difference is smaller than the error bar.
We have amended the CMBFAST code of Seljak \& Zaldarriaga
to compute $\sigma_8$ for \ldcdm models 
with Bunn \& White's (1997) COBE normalization
and find $\sigma_8=1.06$ for a spectral index $n=1$.

Initially, $\Gamma\equiv\Omega_M h=0.31$, which
is consistent at $\sim 1.0\sigma$
with the value of $\Gamma=0.24\pm 0.06$ derived
from large-scale structure observations (Feldman, Kaiser, \& Peacock 1994).
Due to the late decay of density fluctuations,
power will be transferred from small scales
to large scales. 
This effect may render an effective $\Gamma$ at $z=0$
smaller than its initial value, more consistent
with observations.
The universal deceleration parameter
$q_0$ is found to be $-0.38$ (compared to $-0.40$ in the corresponding
$\Lambda$ model),
which is consistent with the observations of high-z SNe (Schmidt \etal 1998).
The essential parameters of the model are summarized in
Table (1), where we also list a high $\Omega_b$ model.

\subsection{More On Small Scales -- Dwarf Galaxies, Low Surface Brightness Galaxies, Dark Galaxies}

In the \ldcdm model
CDM halos become less concentrated only at low redshift.
It is tempting to conjecture that 
CDM dominated galaxies (mostly dwarf galaxies)
at moderate redshift become 
the low surface brightness galaxies 
in the local universe (LSBGs; Bothun \etal 1987; McGaugh 1992).
We suggest that 
the moderate-redshift faint blue compact objects (FBOs;
Koo 1986; Tyson 1988; Cowie \etal 1988)
are CDM dominated 
dwarf galaxies undergoing initial starbursts
at $z\sim 1.0$, 
and subsequently expand
to become the LSBGs
seen today.
There are several pieces of observational data that together provide 
evidence consistent with this scenario.
First, the number density of FBOs is
consistent with local LSBGs (Babul \& Ferguson 1996).
Second, the surface brightness of the LSBGs
is roughly about $1.4$ magnitude (i.e., a factor of $4.4$)
dimmer than 
high surface brightness galaxies (HSBGs; McGaugh 1994),
which is consistent with the expectation 
that FBOs have high surface brightness,
and the expansion of size
by a factor of $\sim 2$ just gives the indicated magnitude difference.
Third, the LSBGs are weakly clustered (Mo, McGaugh, \& Bothun 1994),
in agreement with FBOs (Efstathiou \etal 1991);
since galaxy clustering is not expected to involve
rapidly from $z=1$ to $z=0$ (Katz, Hernquist, \& Weinberg 1999;
Cen \& Ostriker 2000),
the comparison between objects at $z\sim 1$ (FBOs)
and at $z=0$ (LSBGs) is indicative.

While the connection between FBOs and LSBGs
was previously made by McGaugh (1994),
there is a major difference between his proposal and ours.
While he argues that FBOs are LSBGs at high $z$,
we require that FBOs be HSBGs at high $z$,
which subsequently evolve to LSBGs.
There is some indication
that our picture is in better
agreement with recent HST observations (Glazebrook \etal 2000).
Furthermore, our picture of 
moderate-redshift FBOs being HSBGs
would naturally enable 
FBOs to experience necessary initial
starbursts (Lacey \& Silk 1991; Babul \& Rees 1992;
Gardner \etal 1993; Lacey \etal 1993; Kauffmann \etal 1994);
local LSBGs are not starburst galaxies.
Babul \& Ferguson (1996)
have also
made the connection between FBOs and LSBGs
based on detailed modeling of 
the star formation history of FBOs.
The major difference between our model and theirs
is that we do not require gas to be expelled from FBOs.
It is noted
that LSBGs are rather gas rich, {\it not gas poor},
and many LSBGs are fairly massive systems,
for which it seems unlikely that gas can be easily driven out
(Mac Low and Ferrara 1999).


We would go further to propose that there
exist two additional classes of some more numerous, 
(on average) 
lower mass CDM halos: 
ultra-low surface brightness galaxies (ULSBGs; 
Impey \& Bothun 1997),
and ``dark" galaxies (DGs).
ULSBGs have lower star formation rate than LSBGs,
whereas DGs were never given a chance to form a significant amount of
stars and were expanded away to remain dark.
The existence of a sharp drop-off of star formation rate
below a cut-off density (Kennicutt 1998) bodes well for this picture.
Most of the excess number of dwarf halos
found in N-body simulations of the \lcdm model
will now be ULSBGs and DGs in the \ldcdm model.
ULSBGs and DGs as well as LSBGs should be low in metallicity and rich in gas.

Peebles (1999) has repeatedly reminded us
of a potential problem with 
CDM based models:  
Why are there not many dwarf galaxies with escape
velocities greater than $\sim 20$km/sec in the voids,
which are expected to form from low $\sigma$
density peaks at low redshift? 
One answer born out in our picture
is that the universe today may be
filled with numerous LSBGs, ULSBGs, and DGs
in voids (as well as in some other regions).
The fact that low $\sigma$ small-scale density peaks 
are more affected by
the modulation by large-scale waves, which determine
the large-scale structures such as voids and clusters,
implies that low $\sigma$ peaks
virialize later in the voids
than their counterparts in higher density environments;
it may be that FBOs were never formed in voids because of halo expansion.
This effect may have left the voids filled with only ULSBGs
and DGs, as observations 
do not seem to find any significant excess of LSBGs in voids
(Schombert, Pildis, \& Eder 1997).


\subsection{Other Implications of the \ldcdm Model}

1. The fact that the \ldcdm model
has a lower non-relativistic matter density at $z=0$ 
dictates a higher $\sigma_8$
compared to the canonical \lcdm model,
when both are normalized to the
abundance of local clusters of galaxies.
The shape of the power spectrum of the \ldcdm model 
is similar to that of the canonical \lcdm model.
This, combined with a slower 
linear growth factor,
a higher effective $\Gamma$ ($\equiv \Omega_M h$)
and a higher non-relativistic matter density at high redshift
says that \ldcdm model has
more power on all scales at high redshift 
compared to the corresponding \lcdm model.
This may prove to be attractive 
in light of the discovery of very high
redshift, very luminous quasars (Fan \etal 2000).
On larger scales, it appears that the canonical \lcdm model
underpredicts the abundance of high temperature
clusters at $z\sim 0.83$ 
by a factor of about ten (see Figure 1 of Bahcall \& Fan 1998).
The \ldcdm should have more high-z clusters but detailed simulations
are needed to quantify this.

2. Galaxies at high redshift ($z\geq 1.0-2.0$)
is expected to be smaller, on average, than local ones. 
In particular, there should be an excess number of small galaxies
at $z\geq 1.0-2.0$,
which may have already 
been seen in the Hubble Deep Field (e.g., Giallongo \etal 2000).

3. The comoving distance to any redshift is shorter
in the \ldcdm model than in the corresponding \lcdm model,
which 
lessens the constraint on $\lambda_0$ imposed 
by gravitational lensing of galaxies (Kochanek 1996).

4. ULSBGs and DGs may make a sizable contribution to the
number of Lyman alpha clouds and Lyman limit systems 
at low redshift and they are likely to cluster around
large galaxies.
They could provide the source of 
the ``Type 1" population of Lyman alpha clouds at low redshift
as proposed by Bahcall \etal (1996).

5. It is interesting to see whether the decay of CDM particles
has 
consequences on the dynamics of stars
in our own Galaxy.
A star at a galacto-centric
distance $r$ will have an extra outward velocity
$v_r={r\over M} {dM\over dt}$ (this is called ``K-term" in galactic dynamics),
which gives $v_r = 0.4(r/10\kpc)$km/s for
$\tau=1.44t_0$ (assuming CDM
and baryonic mass are equal within the relevant radius),
not inconsistent with observations (e.g., Ovenden \& Byl 1976).

We do not attempt to identify a particle physics model 
for the decaying CDM with the designed properties,
but possible candidates for them
are heavy neutrinos that decay to light neutrinos
plus majoron (Gelmini, Schramm, \& Valle 1984),
or supersymmetric models involving the nonradiative decay of
the gravitino (Olive, Schramm, \& Srednicki 1984).
Finally, CDM decay epoch appears to
present another cosmic coincidence, as does the cosmological
constant.
We would conjecture, as the risk of being overly speculative,
that the two parameters may be related
at a more fundamental level.

\subsection{Tests of the \ldcdm Model}

One clean test that could either falsify
or confirm the \ldcdm model
will come from measuring
the ratio of gas to total mass in
clusters of galaxies, $M_{gas}/M_{tot}$.
We predict that
$M_{gas}/M_{tot}$ should display a trend: lower at higher
redshift. 
The ratio should be lower by $31\%$ at $z=1$ than 
at $z=0$.
This is a potentially observable signature
by X-ray observations (e.g., Jones \& Forman 1992) or observations of 
the Sunyaev-Zel'dovich (SZ) effect (e.g.,
Carlstrom, Joy, \& Grego 1996).

Similarly, the mass-to-light ratio for galaxies, particularly dwarf galaxies,
should show a trend of increasing with redshift.
However, complications are expected in this regard due
to much more involved astrophysical processes
concerning star-formation. 
Nevertheless, 
attempts to try to detect this trend are worthwhile.

\section{Conclusions}

We have shown that, by allowing
for one half of the CDM particles
to decay by $z=0$ 
into relativistic particles,
the problem of excess small-scale power in the \lcdm model is remedied.
In essence,
the decay of CDM particles 
results in reduction of CDM mass
and expansion of the halo, and 
lowers the concentration of CDM in the inner region 
to brings the model
into agreement with observations of dwarf galaxies.
The problem of excess number of dwarf galaxies
in the \lcdm is solved by the same
mechanism but manifested in a quite different way:
instead of suppressing the number of predicted
dwarf halos, 
we argue that these dwarf halos failed to form
a sufficient amount of stars to be identified
as dwarf galaxies even in our local neighborhood.
It is important to search for these gas-rich 
(not necessarilly neutral gas-rich) dark galaxies in the Local Group. 
The model is consistent with COBE, the local abundance of rich
clusters of galaxies, the age constraint and $q_0$ from high-z SNe.

A test of the model will be provided
by measuring the evolution of gas fraction in clusters.
The prediction is that the gas fraction should decrease
with redshift and is smaller by $31\%$ at $z=1$ than
at $z=0$. X-ray and SZ effect observations
should provide such a test.

\acknowledgments
I thank Matias Zaldarriaga for help on CMBFAST code,
Jerry Ostriker, David Spergel, Paul Steinhardt and
Michael Strauss for very useful discussions and comments,
Jim Peebles for encouragement,
Vijay Narayanan for a very careful reading of the manuscript.
This research is supported in part by grants AST-9803137
and ASC-9740300.

\begin{deluxetable}{cccccccccc} 
\tablewidth{0pt}
\tablenum{1}
\tablecolumns{12}
\tablecaption{Two COBE and Cluster-Normalized \ldcdm Models} 
\tablehead{
\colhead{Model} &
\colhead{$\Omega_{0,CDM}$} &
\colhead{$\Omega_{0,b}$} &
\colhead{$\Omega_{0,r}$} &
\colhead{$\lambda_0$} &
\colhead{$h$} &
\colhead{$n$} &
\colhead{$\sigma_8$} &
\colhead{$t_0$ (Gyrs)} &
\colhead{$q_0$}}

\startdata
{\rm fiducial} & $0.234$ & $0.044$ & $0.122$ & $0.60$ & $0.60$ & $1.00$ & $1.06$ & $13.9$  & $-0.38$ \nl 
{\rm high $\Omega_b$} & $0.220$ & $0.066$ & $0.114$ & $0.60$ & $0.60$ & $1.00$ & $0.99$ & $13.9$  & $-0.38$ \nl 
\enddata
\end{deluxetable}

\begin{figure}
\begin{picture}(400,200)
\psfig{figure=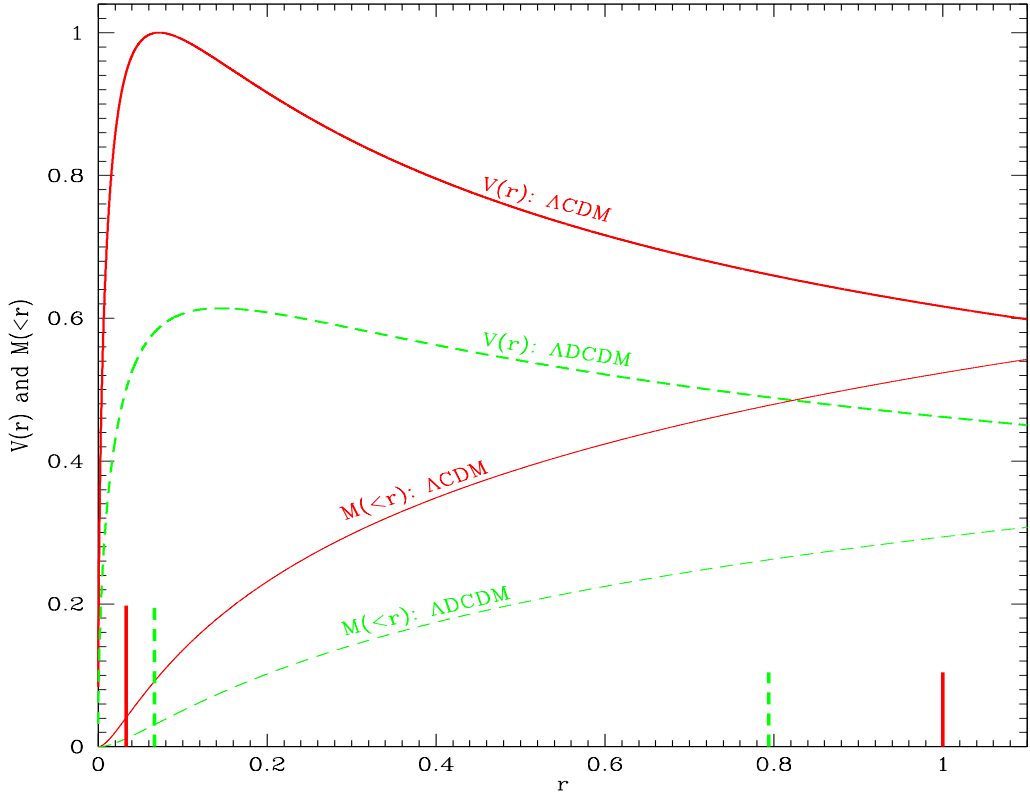,height=12.0cm,width=15.0cm,angle=0.0}
\end{picture}
\caption{
shows the
rotation curves (thick) and mass profiles (thin) 
for the initial (solid; \lcdm) and final (dashed; \ldcdm) halo,
both having the NFW profile but different virial radii 
($r_{200}$).
Both x and y axes have arbitrary units.
The initial concentration of $c_i=30$ is used
for this illustration, although the relative effect 
is quite insensitive to the value of $c_i$.
The long vertical bars indicate the 
core radius for the initial (\lcdm; solid) 
and final (\ldcdm; dashed) halo.
The short vertical bars indicate the 
virial radius ($r_{200}$) for the initial (\lcdm; solid) 
and final (\ldcdm; dashed) halo.
The final radius of maximum rotation velocity
is larger than the initial radius of maximum rotation velocity
by a factor of $2$.
The rotation velocity at virial radius is reduced
by about $20\%$ ($V_{200,f}\sim 0.8 V_{200,i}$)
and the maximum rotation velocity reduced
by about $40\%$ ($V_{max,f}\sim 0.6 V_{max,i}$).
The mass of a halo 
within a fixed proper radius in the \ldcdm
model is reduced by a factor of $\sim 8.0$ in the inner
region and a factor of $\sim 1.6$ in outer region.
}
\end{figure}

\end{document}